\documentclass[prc,twocolumn,aps,superscriptaddress,showpacs]{revtex4}
\usepackage{amssymb}
\usepackage{amsmath,bm}
\usepackage{graphicx}
\usepackage[normalem]{ulem}
\usepackage{booktabs}
\usepackage{threeparttable}
\usepackage[dvips]{color}
\usepackage{leftidx}

\renewcommand\sout{\bgroup \color{red} \ULdepth=-.5ex \ULset}

\begin{document}

\title{Empirical information on nuclear matter fourth-order symmetry energy from an extended nuclear mass formula}
\author{Rui Wang}
\affiliation{School of Physics and Astronomy and Shanghai Key Laboratory for Particle
Physics and Cosmology, Shanghai Jiao Tong University, Shanghai 200240, China}
\author{Lie-Wen Chen\footnote{Corresponding author (email: lwchen$@$sjtu.edu.cn)}}
\affiliation{School of Physics and Astronomy and Shanghai Key Laboratory for Particle
Physics and Cosmology, Shanghai Jiao Tong University, Shanghai 200240, China}
\affiliation{Center of Theoretical Nuclear Physics, National Laboratory of Heavy Ion
Accelerator, Lanzhou 730000, China}
\date{\today}

\begin{abstract}

We establish a relation between the equation of state of nuclear matter and the fourth-order symmetry energy $a_{\rm{sym,4}}(A)$ of finite nuclei in a semi-empirical nuclear mass formula by self-consistently considering the bulk, surface and Coulomb contributions to the nuclear mass.
Such a relation allows us to extract information on nuclear matter fourth-order symmetry energy $E_{\rm{sym,4}}(\rho_0)$ at normal nuclear density $\rho_0$ from analyzing nuclear mass data. Based on the recent precise extraction of $a_{\rm{sym,4}}(A)$ via the double difference of the ``experimental'' symmetry energy extracted from nuclear masses, for the first time, we estimate a value of $E_{\rm{sym,4}}(\rho_0) = 20.0\pm4.6$ MeV.
Such a value of $E_{\rm{sym,4}}(\rho_0)$ is significantly larger than the predictions from mean-field models and thus suggests the importance of considering the effects of beyond the mean-field approximation in nuclear matter calculations.

\end{abstract}


\maketitle

\section{Introduction}
The determination of the isospin dependent part of nuclear matter equation of state (EOS) has become a hot topic in both nuclear physics and astrophysics during the last decades~\cite{LiBA98,Dan02,Lat04,Ste05,Bar05,LCK08,Tra12,Hor14,LiBA14,Bal16,Oer17,LiBA17}.
The nuclear matter EOS tells us its energy per nucleon $E(\rho, \delta)$ as a function of density $\rho$ $=$ $\rho_{\rm{n}}$ $+$ $\rho_{\rm{p}}$ and isospin asymmetry $\delta$ $=$ $(\rho_{\rm{n}} - \rho_{\rm{p}})/\rho$ with $\rho_{\rm{n}}$ ($\rho_{\rm{p}}$) being the neutron (proton) density.
The parabolic approximation to nuclear matter EOS, i.e., $E(\rho, \delta)$ $\approx $ $E(\rho,\delta=0)$ $+$ $E_{\rm{sym}}(\rho)\delta^2$, is adopted widely with the symmetry energy defined as $E_{\rm{sym}}(\rho)=$ $\frac{1}{2!}\frac{\partial^2 E(\rho,\delta)}{\partial\delta^2}\big|_{\delta = 0}$.
The feasibility of the parabolic approximation is practically justified in various aspects of nuclear physics, especially in finite nuclei where the $\delta^2$ value is usually significantly less than one.
Nevertheless, in neutron stars where the $\delta$ could be close to one, a sizable higher-order terms of isospin dependent part of nuclear matter EOS, e.g., the term $E_{\rm{sym,4}}(\rho)\delta^4$ with the fourth-order symmetry energy defined as $E_{\rm{sym,4}}(\rho)=$ $\frac{1}{4!}\frac{\partial^4 E(\rho,\delta)}{\partial\delta^4}\big|_{\delta = 0}$, may have substantial effects on the properties such as the proton fraction at beta-equilibrium, the core-crust transition density and the critical density for the direct URCA process~\cite{Zha01,Ste06,Xu09,Cai12,Sei14}.

To the best of our knowledge, unfortunately,
there is so far essentially no experimental information on the magnitude of $E_{\rm{sym,4}}(\rho)$, even at normal nuclear density $\rho_0$.
Theoretically, the mean-field models generally predict the magnitude of $E_{\rm{sym,4}}(\rho_0)$ is less than $2$ MeV~\cite{ChenLW09,Cai12,Arg17,PuJ17}. A value of $E_{\rm{sym,4}}(\rho_0)=1.5$ MeV is obtained from chiral pion-nucleon dynamics~\cite{Kai15}. The recent study~\cite{Nan16} within the quantum molecular dynamics (QMD) model indicates that the $E_{\rm{sym,4}}(\rho_0)$ could be as large as $3.27 \sim 12.7$ MeV depending on the interactions used. Based on an interacting Fermi gas model, a significant value of $7.18\pm2.52$ MeV~\cite{Cai15} is predicted for the kinetic part of $E_{\rm{sym,4}}(\rho_0)$ by considering the high-momentum tail~\cite{Hen14} in the single-nucleon momentum distributions that could be due to short-range correlations of nucleon-nucleon interactions. In addition, the divergence of the isospin-asymmetry expansion of nuclear matter EOS in many-body perturbation theory is discussed in Refs.~\cite{Kai15,Wel16}.
Therefore, the magnitude of $E_{\rm{sym,4}}(\rho_0)$ is currently largely uncertain and it is of critical importance to obtain some experimental or empirical information on $E_{\rm{sym,4}}(\rho_0)$.

Conventionally nuclear matter EOS is quantitatively characterized in terms of a few characteristic coefficients through Taylor expansion in density at $\rho_0$, e.g., $E(\rho, \delta = 0)$ $=$ $E_0(\rho_0)$ $+$ $\frac{1}{2!}K_0\chi^2$ $+$ $\frac{1}{3!}J_0\chi^3$ $+$ ${\cal O}(\chi^4)$ and $E_{\rm{sym}}(\rho)$ $=$ $E_{\rm sym}(\rho_0)$ $+$ $L\chi$ $+$ $\frac{1}{2!}K_{\rm{sym}}\chi^2$ $+$ ${\cal O}(\chi^3)$ with $\chi$ $=$ $\frac{\rho - \rho_0}{3\rho_0}$.
The density in the interior of heavy nuclei is believed to nicely approximate to saturation density of symmetric nuclear matter (nuclear normal density) $\rho_0$ and the empirical value of $\rho_0\approx$ $0.16$ fm$^{-3}$ has been obtained from measurements on electron or nucleon scattering off heavy nuclei~\cite{Jac74}.
Our knowledge on nuclear matter EOS largely stems from nuclear masses based on various nuclear mass formulae.
By analyzing the data on nuclear masses with various nuclear mass formulae (see, e.g., Ref.~\cite{Mol12}), consensus has been reached on $E_0(\rho_0)$ and $E_{\rm sym}(\rho_0)$ with $E_0(\rho_0)$ $\approx$ $-16.0~\rm MeV$ and $E_{\rm sym}(\rho_0)$ $\approx$ $32.0~\rm MeV$.
These empirical values on $E_0(\rho_0)$ and $E_{\rm sym}(\rho_0)$ are of critical importance for our understanding on nuclear matter EOS.

Generally speaking, it is very hard to determine the higher-order parameter $E_{\rm{sym,4}}(\rho_0)$ and the fourth-order symmetry energy $a_{\rm sym,4}(A)$ of finite nuclei from simply fitting nuclear masses within nuclear mass formulae since the term $a_{\rm sym,4}(A)I^4$ ($I =\frac{N-Z}{A}$ with $N$ and $Z$ being the neutron and proton number, respectively, and $A=$ $N$ $+$ $Z$ is mass number) is considerably small compared to other lower-order terms in the mass formula for known nuclei, even for the predicted dripline nuclei~\cite{WangR15}.
Recently, however, by approximating $a_{\rm sym,4}(A)$ to a constant $c_{\rm sym,4}$ in the mass formula, several studies~\cite{Jia14,Jia15,Wan15,Tia16} have been performed to extract $c_{\rm sym,4}$ from analyzing the double difference of the ``experimental'' symmetry energy extracted from nuclear mass data, and robust results with high precision have been obtained, i.e., a sizable positive value of $c_{\rm sym,4}=$ $3.28\pm0.50~\rm MeV$ or $8.47\pm0.49$ MeV is obtained in Ref.~\cite{Jia14}, depending on the Wigner term form in the mass formula. More recently, a value of $c_{\rm sym,4}=$ $8.33\pm1.21~\rm MeV$ is extracted in Ref.~\cite{Tia16} using similar analysis on nuclear masses. These results provide the possibility to extract information on $E_{\rm{sym,4}}(\rho_0)$.

In this work, by self-consistently considering the bulk, surface and Coulomb contributions to the nuclear mass, we extend the mass formula of Ref.~\cite{Dan03} to additionally include the corrections due to central density variation of finite nuclei and the higher-order fourth-order symmetry energy term $a_{\rm{sym,4}}(A)I^4$.
In this extended mass formula, a explicit relation between $a_{\rm{sym,4}}(A)$ and $E_{\rm{sym,4}}(\rho_0)$ is obtained. We demonstrate for the first time that the precise value of $c_{\rm sym,4}$ obtained recently from nuclear mass analysis allows us to estimate a value of $E_{\rm{sym,4}}(\rho_0) = 20.0\pm4.6$ MeV.

\section{Nuclear mass formula}
There have been a number of nuclear mass models which aim to describe the experimental nuclear mass database and predict the mass of unknown nuclei.
Nowadays, some sophisticated mass formulae~\cite{Roy08,Mol12,Wan10,Wan14}~(with shell and pairing corrections) can reproduce the measured masses of more than $2000$ nuclei with a root-mean-square deviation of merely several hundred $\rm{keV}$s.
These mass formulae provide us empirical information about the EOS of nuclear matter, especially its lower-order characteristic parameters $E_0(\rho_0)$, $E_{\rm sym}(\rho_0)$ and so forth.

To relate the coefficients in the mass formula to the EOS of nuclear matter, one can express the binding energy $B(N,Z)$ of a nucleus with $N$ neutrons and $Z$ protons in terms of the bulk energy of nuclear matter in the interior of the nucleus plus surface corrections and Coulomb energy.
Based on such an argument, Danielewicz~\cite{Dan03} developed a mass formula with a self-consistent $A$-dependent symmetry energy $a_{\rm sym}(A)$ of finite nuclei.
Considering that the central density $\rho_{\rm cen}$ in nuclei generally depends on $N$ and $Z$ and deviates from $\rho_0$, we here extend the mass formula of Ref.~\cite{Dan03} by considering the deviation of $\rho_{\rm cen}$ from $\rho_0$, and additionally including the higher-order $I^4$ terms.
In such a framework, a nucleus with $N$ neutrons and $Z$ protons is assumed to localize inside an effective sharp radius $R$, i.e.,
\begin{equation}
R = r_0\big[1 + 3\chi_{\rm{cen}}(N,Z)\big]^{-1/3}A^{1/3},
\label{radius}
\end{equation}
where $r_0$ is a constant satisfying $\frac{4}{3}\pi\rho_0r_0^3 = 1$ and $\chi_{\rm cen}$ $=$ $(\rho_{\rm cen} - \rho_0)/3\rho_0$ is a dimensionless variable characterizing the deviation of $\rho_{\rm cen}$ from $\rho_0$.
Furthermore, we denote the volume~(surface) neutron excess as $\Delta_{\rm v}$ $=$ $N_{\rm v} - Z_{\rm v}$~($\Delta_{\rm s}$ $=$ $N_{\rm s} - Z_{\rm s}$), where $N_{\rm v}$ ($Z_{\rm v}$) and $N_{\rm s}$ ($Z_{\rm s}$) represent the neutron (proton) number in the volume and surface regions of the nucleus, respectively, with $N_{\rm v}$ $+$ $N_{\rm s}$ $=$ $N$ and $Z_{\rm v}$ $+$ $Z_{\rm s}$ $=$ $Z$.
Generally, $\chi_{\rm cen}$ and $\Delta_{\rm v}$~($\Delta_{\rm s}$) depend on $N$ and $Z$ of the nucleus and can be determined from equilibrium conditions, and this is consistent with the argument of the droplet model (see, e.g., Ref.~\cite{Rei06}).

In the present work, the nuclear binding energy consists of volume term $B_{\rm v}$, surface term $B_{\rm s}$ and Coulomb term $B_{\rm c}$.
The volume part of the binding energy can be treated in nuclear matter approximation, i.e.,
\begin{equation}
\begin{split}
B_{\rm v} & \approx A\Big[E_0(\rho_0) + \frac{1}{2}K_0\chi_{\rm cen}^2 + E_{\rm sym}(\rho_0)\big(\frac{\Delta_{\rm v}}{A}\big)^2\\
 & + L\chi_{\rm cen}\big(\frac{\Delta_{\rm v}}{A}\big)^2 + E_{\rm sym,4}(\rho_0)\big(\frac{\Delta_{\rm v}}{A}\big)^4\Big].
\label{BE-V}
\end{split}
\end{equation}
The surface term comes from surface tension and symmetry potential~(detailed argument can be found in Ref.~\cite{Dan03}), and it can be expressed as
\begin{eqnarray}
B_{\rm s} & = & \Big[\sigma_0 - \sigma_{\rm I} \big(\frac{\Delta_{\rm s}}{S}\big)^2\Big]4\pi R^2 + \frac{2\sigma_{\rm I}}{4\pi R^2}\Delta_{\rm s}^2\nonumber \\
& \approx & E_{\rm s0}(1-2\chi_{\rm cen})A^{\frac{2}{3}} + \beta(1+2\chi_{\rm cen})A^{\frac{4}{3}}\big(\frac{\Delta_{\rm s}}{A}\big)^2,
\label{BE-S}
\end{eqnarray}
where $\sigma_0$ ($\sigma_{\rm I}$) represents the isospin independent (dependent) surface tension, $S = 4\pi R^2$ is the surface area of the nucleus, and we define $E_{\rm s0} = 4\pi r_0^2\sigma_0$ and $\beta = \frac{\sigma_{\rm I}}{4\pi r_0^2}$. Eq.~(\ref{radius}) has been used to obtain the second line in Eq.~(\ref{BE-S}).
For Coulomb energy, for simplicity we adopt the following simple form without exchange term, i.e.,
\begin{equation}
B_{\rm c} = \frac{3}{5}\frac{e^2}{4 \pi \epsilon_0}\frac{1}{R}Z^2 \approx a_{\rm c} A^{-1/3}Z^2(1 + \chi_{\rm cen}),
\label{BE-C}
\end{equation}
with $a_c=\frac{3}{5}\frac{e^2}{4\pi\epsilon_0r_0}$.

The equilibrium condition of nuclei can be obtained from variations of the binding energy $B(N,Z)$ of the nucleus with respect to $\chi_{\rm cen}$ and $\Delta_{\rm v}$, i.e.,
\begin{equation}
\frac{\partial B(N,Z)}{\partial\chi_{\rm cen}} = 0, \qquad \frac{\partial B(N,Z)}{\partial\Delta_{\rm v}} = 0,
\label{Variation}
\end{equation}
from which we can obtain $\chi_{\rm cen}$ and $\Delta_{\rm v}$~($\Delta_{\rm s}$) for different $A$ and $Z$.
The first equation means the mechanical equilibrium and tells us how the surface energy, Coulomb energy and the isospin dependent part of volume energy affect the value of $\rho_{\rm cen}$, while the second equation represents the balance of the isospin asymmetry chemical potential between the volume and surface regions.

To solve Eq.~(\ref{Variation}), we expand $\chi_{\rm cen}$ in terms of $\frac{\Delta_{\rm v}}{A}$, and then expand $(\frac{\Delta_{\rm v}}{A})^2$ in terms of $I$, i.e.,
\begin{eqnarray}
\chi_{\rm cen} & =& \chi_0 + \chi_2 \big(\frac{\Delta_{\rm v}}{A}\big)^2 + {\cal O}\Big[\big(\frac{\Delta_{\rm v}}{A}\big)^4\Big],
 \label{chi}\\
\big(\frac{\Delta_{\rm v}}{A}\big)^2  & = &  D_2I^2 + {\cal O}(I^4),
\label{Deltav}
\end{eqnarray}
where the expansion coefficients $\chi_0$, $\chi_2$ and $D_2$ might depend on $A$ or $Z$, consistent with calculations from the droplet model~\cite{Rei06} and the Thomas-Fermi approximation~\cite{WangR17}.
Using Eqs.~(\ref{BE-V}), (\ref{BE-S}) and (\ref{BE-C}) and substituting $B(N,Z)$ $=$ $B_{\rm v}$ $+$ $B_{\rm s}$ $+$ $B_{\rm c}$ into Eq.~(\ref{Variation}) leads to the following two equations
\begin{widetext}
\begin{eqnarray}
\frac{\partial B}{\partial\chi_{\rm cen}} & = & A \Big[K_0 \chi_{\rm cen} + E_{\rm sym}(\rho_0)\big(\frac{\Delta_{\rm v}}{A}\big)^2\Big] - 2 E_{\rm s0} A^{\frac{2}{3}} + 2 \beta A^{\frac{4}{3}} \big(\frac{\Delta_{\rm s}}{A}\big)^2 + a_{\rm c} Z^2 A^{-\frac{1}{3}} = 0\label{vrchi},\\
\frac{\partial B}{\partial\Delta_{\rm v}} & = & 2\big(E_{\rm sym}(\rho_0) + L \chi_{\rm cen}\big)\frac{\Delta_{\rm v}}{A} + 4 E_{\rm sym,4}(\rho_0)\big(\frac{\Delta_{\rm v}}{A}\big)^3 - 2\beta A^{\frac{1}{3}}\frac{\Delta_{\rm s}}{A} (1 + 2 \chi_{\rm cen}) = 0\label{vrdlt}.
\end{eqnarray}
\end{widetext}
By eliminating $\frac{\Delta_{\rm s}}{A}$ in Eq.~(\ref{vrchi}) and Eq.~(\ref{vrdlt}), we obtain
\begin{equation}
\begin{split}
 & K_0 A \chi_{\rm cen} - 2 E_{\rm s0} A^{\frac{2}{3}} + a_{\rm c} Z^2 A^{-\frac{1}{3}} + {\cal O}(A^{\frac{1}{3}})\\
 + & \big[LA + {\cal O}(A^{\frac{2}{3}})\big] \big(\frac{\Delta_{\rm v}}{A}\big)^2 + {\cal O}\Big[\big(\frac{\Delta_{\rm v}}{A}\big)^4\Big] = 0.
 \label{vrpls}
\end{split}
\end{equation}
From Eqs.~(\ref{vrpls}) and (\ref{chi}), one can obtain $\chi_0$ and $\chi_2$, i.e.,
\begin{equation}
\chi_0(A,Z) = \frac{2E_{\rm s0}A^{2/3} - a_{\rm c}Z^2A^{-1/3}}{AK_0},
\label{chi0}
\end{equation}
which represents the modification of the central density of finite nuclei due to the surface and Coulomb energy, and
\begin{equation}
\chi_2 = -\frac{L}{K_0},
\label{chi2}
\end{equation}
which determines the modification of nuclear central density due to the isospin dependent part of nuclear matter EOS.
On the other hand, one can figure out the relation between $(\frac{\Delta_{\rm v}}{A})^2$ and $I$ from Eq.~(\ref{vrdlt}), i.e.,
\begin{eqnarray}
I^2 & = & \big(\frac{\Delta_{\rm v}}{A} + \frac{\Delta_{\rm s}}{A}\big)^2 \nonumber \\
& \approx & \Big[\frac{\big(E_{\rm sym}(\rho_0) + \beta A^{\frac{1}{3}}\big)^2}{\beta^2A^{2/3}}  + O(A^{\frac{1}{3}})\Big]\big(\frac{\Delta_{\rm v}}{A}\big)^2,
\end{eqnarray}
from which, together with Eq.~(\ref{Deltav}), one then obtains
\begin{equation}
D_2(A) = \frac{1}{(1 + \frac{E_{\rm sym}(\rho_0)}{\beta}A^{-1/3})^2}.
\label{D2}
\end{equation}
As can be seen from Eq.~(\ref{Deltav}), the $D_2(A)$ reflects the fraction of volume neutron excess in the total neutron excess of a nucleus with mass number $A$. The ratio $E_{\rm sym}(\rho_0)/\beta$ is the so-called symmetry volume-surface ratio~\cite{Dan03}, which can be considered as an independent parameter.

From Eqs.~(\ref{chi0}), (\ref{chi2}) and (\ref{D2}), $\chi_{\rm cen}$ and $\Delta_{\rm v}$ in Eqs.~(\ref{chi}) and (\ref{Deltav}) can be expressed in terms of $A$ and $Z$, and thus the nuclear binding energy $B(N,Z)$ $=$ $B_{\rm v}$ $+$ $B_{\rm s}$ $+$ $B_{\rm c}$ can be recast into
\begin{equation}
\begin{split}
 B(A,Z) & =A\times \big[c_{00}(A,Z)  + c_{01}(A,Z)A^{-\frac{1}{3}}\\
 & + a_{\rm{sym}}(A,Z)I^2 + a_c(1+\chi_0)Z^2A^{-\frac{4}{3}}\\
 & + a_{\rm{sym,4}}(A)I^4\big],
\end{split}
\label{MF}
\end{equation}
where $c_{00}$ and $c_{01}$ characterize the isospin independent parts of $ B(A,Z)$ with
\begin{eqnarray}
c_{00}(A,Z) & = & E_0(\rho_0) + \frac{1}{2}K_0\chi_0^2(A,Z),\\
c_{01}(A,Z) & = & E_{\rm{s}0}\big[1 - 2\chi_0(A,Z)\big],
\end{eqnarray}
while the symmetry energy $a_{\rm sym}(A,Z)$ and the fourth-order symmetry energy $a_{\rm{sym,4}}(A)$ of finite nuclei can be expressed, respectively, as
\begin{eqnarray}
a_{\rm{sym}}(A,Z) & = & D_2(A)\Big[E_{\rm sym}(\rho_0) + a_{\rm c}Z^2\chi_2A^{-\frac{4}{3}}\nonumber\\
 & + & \Big(\frac{E_{\rm sym}^2(\rho_0)}{\beta}-2E_{\rm{s}0}\chi_2\Big)A^{-\frac{1}{3}}\Big]\label{asym},\\
a_{\rm{sym,4}}(A) & = & D_2^2(A)\Big(E_{\rm sym,4}(\rho_0)-\frac{L^2}{2K_0}\Big)\label{asym4}.
\end{eqnarray}
Noting that $a_{\rm{sym}}(A,Z)$ in Eq.~(\ref{asym}) includes a small $Z$-dependent term $a_{\rm{c}}A^{-4/3}Z^2\chi_2D_2$, which comes from the modification of $\rho_{\rm{cen}}$ due to the Coulomb energy.

The mass formula Eq.~(\ref{MF}) is an extended form of the mass formula of Ref.~\cite{Dan03} and the latter can be obtained from the former by omitting the corrections due to the central density variation of finite nuclei (i.e., setting $\chi_0=$ $\chi_2 =0$) and the higher-order $I^4$ term.
In particular, by setting $\chi_2$ $=$ $0$, $a_{\rm{sym}}(A,Z)$ is then reduced to a simpler form of $a_{\rm sym}(A)$~\cite{Dan03}, i.e.,
\begin{equation}
 a_{\rm sym}(A) = E_{\rm sym}(\rho_0)\Big/\Big(1 + \frac{E_{\rm sym}(\rho_0)}{\beta}A^{-\frac{1}{3}}\Big).
\label{asym-3}
\end{equation}
In addition, in the limit of $A$ $\rightarrow$ $\infty$, both the $\chi_0$ (see Eq.~(\ref{chi0})) and the $D_2$ (see Eq.~(\ref{D2})) become to zero, Eq.~(\ref{MF}) is then reduced to the binding energy per nucleon of asymmetric nuclear matter at saturation point where the $B(A,Z)/A$ reaches its minimum value, i.e.,
\begin{equation}
\begin{split}
 E_{\rm sat}(\delta) & = B(A,Z)/A = E_0(\rho_0) + E_{\rm sym}(\rho_0)\delta^2\\
 & + \Big(E_{\rm sym,4}(\rho_0) - \frac{L^2}{2K_0}\Big)\delta^4 + {\cal O}(\delta^6),
\end{split}
\label{Esat}
\end{equation}
In the above, the Coulomb interaction is neglected and the $I$ is replaced by $\delta$. It is seen that Eq.~(\ref{Esat}) is exactly the same as the expression obtained in Ref.~\cite{ChenLW09} for the binding energy per nucleon of asymmetric nuclear matter at saturation point.

It should be pointed out that the $A$-dependence of $a_{\rm sym}(A,Z)$ and $a_{\rm sym,4}(A)$ mainly comes from $D_2(A)$ in Eq.~(\ref{D2}), and the same form of $D_2(A)$ has been obtained in Ref.~\cite{Dan03}.
In nuclear mass formula, the nuclear binding energy per nucleon is usually expanded in two small quantities $A^{-1/3}$ and $I^2$~\cite{Mye69,Mye96} with the coefficient of each term being determined by fitting nuclear masses with optimization.
However, since the value of $\frac{E_{\rm sym}(\rho_0)}{\beta}$ is around $2.4$~\cite{Dan03,Dan17}, $\frac{E_{\rm sym}(\rho_0)}{\beta}A^{-1/3}$ in Eq.~(\ref{D2}) is thus not small enough to obtain a rapid converging $A^{-1/3}$ expansion.
Several studies indicate the convergence of $A^{-1/3}$ expansion is not as good as that of $I^2$ (see, e.g., Ref.~\cite{Rei06}).
For example, the value of $a_{\rm sym}(A)$ (neglecting the small $Z$-dependent term) in the range of known nuclei is found to be very different from its asymptotic value at infinite $A$~(or $a_{\rm sym}(\infty)$ in some literatures)~\cite{Cen09,Liu10,ChenLW11}.
Considering the slow convergence of the $A^{-1/3}$ expansion, in our mass formula we do not expand $a_{\rm sym}(A,Z)$ and $a_{\rm sym,4}(A)$ in terms of $A^{-1/3}$ and exactly retain their $A$ dependence.

We would like to point out when the nuclear binding energy $B(N,Z)$ is expanded in terms of $A^{-1/3}$ and $I^2$, the obtained expressions for the coefficients of the five leading-order terms (i.e., $A$, $A^{\frac{2}{3}}$, $A \times I^2$, $A^{\frac{2}{3}} \times I^2$, $A \times I^4$) in the mass formula Eq.~(\ref{MF}) are complete and self-contained, which means all other characteristic parameters of nuclear matter EOS and surface energy that are not show up in the mass formula (e.g., the higher-order $J_0$ and $K_{\rm{sym}}$) will not contribute to the coefficients of the five leading-order terms.

\section{Symmetry energy of finite nuclei and $E_{\rm sym}(\rho)$}
When the semi-empirical mass formula was first introduced, the symmetry energy term has the simple form of $c_{\rm sym}(N-Z)^2/A$, with a constant symmetry energy coefficient $c_{\rm sym}$.
By fitting the newly released nuclear mass table AME2012~\cite{AME2012}~(all nuclei with $A$ $>$ 20 are considered) with the following simple Bethe-Weizs$\ddot{\rm a}$cker mass formula, i.e.,
\begin{equation}
\begin{split}
 B(N,Z) & = c_{\rm vol}A + c_{\rm sur}A^{2/3} + c_{\rm cou}\frac{Z^2(1 - Z^{-2/3})}{A^{1/3}}\\
 & + c_{\rm sym}\frac{(N-Z)^2}{A} + c_{\rm p}\frac{(-1)^{N} + (-1)^Z}{A^{2/3}},
\end{split}
\label{BW}
\end{equation}
we obtain the constant symmetry energy coefficient $c_{\rm sym} = 22.2$ MeV, the surface coefficient $c_{\rm sur} = 17.33$ MeV and the Coulomb coefficient $c_{\rm cou} = 0.709$ MeV.
Here the constant $c_{\rm sym}$ is just a parameter in mass formula and cannot be simply considered as $E_{\rm sym}(\rho_0)$ in the EOS of nuclear matter since the symmetry energy coefficient $a_{\rm sym}(A,Z)$ in the mass formula is sensitively dependent of the mass number $A$ in the mass region of known nuclei as shown in Eq.~(\ref{asym}).

It is constructive to figure out the relation between $c_{\rm sym}$ and $a_{\rm sym}(A,Z)$.
Since each nucleus was considered equally in our simple fitting, the $c_{\rm sym}$ can be treated as arithmetic average of $a_{\rm sym}(A,Z)$, i.e.,
\begin{equation}
 c_{\rm sym} \approx \langle a_{\rm sym}\rangle = \sum\frac{1}{N_{\rm MN}}a_{\rm{sym}}(A,Z)
\label{asymb}
\end{equation}
where $N_{\rm MN}$ $=$ $2348$ is the number of measured nuclei we used in our simple fitting~(i.e., the measured nuclei in AME2012 with 20 $<$ $A$ $<$ $270$) and the sum runs over all these nuclei.
Substituting Eq.~(\ref{asym}) into Eq.~(\ref{asymb}), one can then obtain
\begin{eqnarray}
 c_{\rm sym} & = & E_{\rm sym}(\rho_0)\sum\frac{1}{N_{\rm MN}}D_2(A)\nonumber\\
  & - & \frac{a_{\rm c}L}{K_0}\sum\frac{1}{N_{\rm MN}}Z^2D_2(A)A^{-\frac{4}{3}}\nonumber\\
  & + & \Big(\frac{E_{\rm sym}^2(\rho_0)}{\beta} +\frac{2E_{\rm s0}L}{K_0}\Big)\sum\frac{1}{N_{\rm MN}}D_2(A)A^{-\frac{1}{3}}\nonumber\\
 & = & E_{\rm sym}(\rho_0)\langle D_2(A)\rangle -\frac{a_{\rm c}L}{K_0}\langle Z^2D_2(A)A^{-\frac{4}{3}}\rangle\nonumber\\
 & + & \Big(\frac{E_{\rm sym}^2(\rho_0)}{\beta} +\frac{2E_{\rm s0}L}{K_0}\Big)\langle D_2(A)A^{-\frac{1}{3}}\rangle\label{asymb2},
\end{eqnarray}
where the summations are the same as that in Eq.~(\ref{asymb}) and the average $\langle X(A,Z)\rangle$ is defined as $\sum X(A,Z)/N_{\rm MN}$.
Similarly, one can obtain the following relations
\begin{eqnarray}
 c_{\rm sur} & \approx & E_{\rm s0}(1 - 2\langle\chi_0(A,Z)\rangle)\nonumber\\
 & = & E_{\rm s0}\Big[1 - 2\Big(\frac{E_{\rm s0}\langle2A^{-\frac{1}{3}}\rangle - a_{\rm c}\langle Z^2A^{-\frac{4}{3}}\rangle}{K_0}\Big)\Big]\label{Es}
\end{eqnarray}
and
\begin{eqnarray}
 c_{\rm cou} & \approx & a_{\rm c}(1+\langle\chi_0(A,Z)\rangle)\nonumber\\
 & = & a_{\rm c}\Big[1 + \Big(\frac{E_{\rm s0}\langle2A^{-\frac{1}{3}}\rangle - a_{\rm c}\langle Z^2A^{-\frac{4}{3}}\rangle}{K_0}\Big)\Big]\label{ac}.
\end{eqnarray}

Using $E_{\rm sym}(\rho_0)/\beta$ $=$ $2.4$~\cite{Dan03,Tia16,Dan17}, we find $\langle D_2(A)\rangle = 0.45$, $\langle D_2(A)A^{-\frac{1}{3}}\rangle = 0.09$ and $\langle Z^2D_2(A)A^{-\frac{4}{3}}\rangle = 2.14$.
Furthermore, from Eqs.~(\ref{Es}) and (\ref{ac}) together with $c_{\rm sur}=17.33$ MeV and $c_{\rm cou}=0.709$ MeV obtained by the simple fitting as well as $K_0$ $=$ $240$ MeV~\cite{Shl06}, we find $E_{\rm s0}$ $=$ $17.96~\rm MeV$ and $a_{\rm c}$ $=$ $0.70~\rm MeV$.
Combining these values of $E_{\rm s0}$ and $a_{\rm c}$ with the empirical value of $E_{\rm sym}(\rho_0)$ $=$ $32.0~\rm{MeV}$, $E_{\rm sym}(\rho_0)/\beta$ $=$ $2.4$, $K_0$ $=$ $240~\rm MeV$ and $L$ $=$ $45.2~\rm MeV$~\cite{ZhangZ13}, we finally obtain $c_{\rm sym}$ $\approx$ $21.8~\rm MeV$, which is in good agreement with the value $22.2$ MeV obtained by simple fitting. Our results also indicate that the $E_{\rm s0}$ and $a_{\rm c}$ can be nicely approximated, respectively, by $c_{\rm sur}$ and $c_{\rm cou}$ in the simple Bethe-Weizs$\ddot{\rm a}$cker mass formula.
The above demonstration suggests that information on $E_{\rm sym}(\rho_0)$ can be extracted inversely from the constant $c_{\rm sym}$.
This feature is rather valuable in the case of $a_{\rm sym,4}$ and it provides an approach to extract $E_{\rm sym,4}(\rho_0)$ through the obtained constant fourth-order symmetry energy coefficient $c_{\rm sym,4}$ in the mass formula.
\label{S-asym}

\section{Fourth-order symmetry energy of finite nuclei and $E_{\rm sym,4}(\rho)$}

\begin{figure}[!htb]
\includegraphics[width=8.5cm]{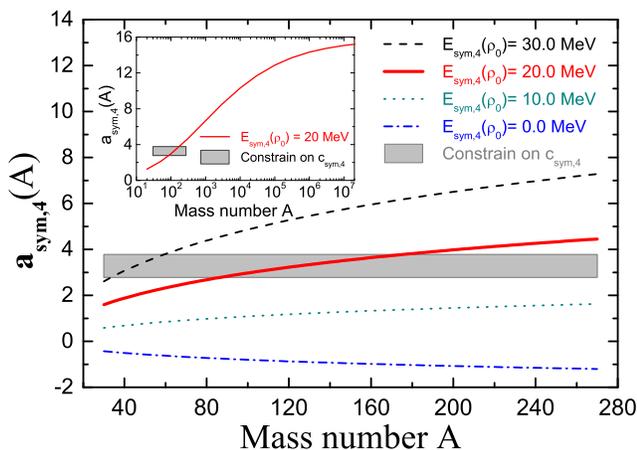}
\caption{The fourth-order symmetry energy $a_{\rm sym,4}(A)$ of finite nuclei as a function of mass number $A$ with $E_{\rm sym,4}(\rho_0)=$ $0.0$ MeV, $10.0$ MeV, $20.0$ MeV and $30.0$ MeV. The gray band represents the constraint of $c_{\rm sym,4}=3.28\pm0.50$ MeV obtained in Ref.~\cite{Jia14}.
The inset shows the asymptotic behavior of $a_{\rm sym,4}(A)$ as a function of $A$ for $E_{\rm sym,4}(\rho_0) = 20.0$ MeV.}
\label{F-asym4}
\end{figure}

Using the empirical values of $K_0 = 240$ MeV, $E_{\rm sym}(\rho_0)/\beta = 2.4$ and $L = 45.2$ MeV, we show in Fig.~\ref{F-asym4} the fourth-order symmetry energy $a_{\rm sym,4}(A)$ of finite nuclei as a function of mass number $A$ for $E_{\rm sym,4}(\rho_0) = $ $0.0$ MeV, $10.0$ MeV, $20.0$ MeV and $30.0$ MeV. Also included in Fig.~\ref{F-asym4} is the constant fourth-order symmetry energy coefficient $c_{\rm sym,4}=3.28\pm0.50$ MeV obtained in Ref.~\cite{Jia14}.
The inset of Fig.~\ref{F-asym4} displays the asymptotic behavior of the obtained $a_{\rm sym,4}(A)$ as a function of the mass number $A$ for $E_{\rm sym,4}(\rho_0) = 20.0$ MeV.
As shown in the inset, the value of $a_{\rm sym,4}(A)$ in the region of $A$ $<$ $270$ is only about one fourth of $a_{\rm sym,4}(\infty)$, indicating the very slow convergence of the $A^{-1/3}$ expansion for $a_{\rm sym,4}(A)$.

Similarly as in our simple fitting of $c_{\rm sym}$ in section~\ref{S-asym}, when extracting the constant fourth-order symmetry energy coefficient $c_{\rm sym,4}$ from analyzing nuclear masses in Ref.~\cite{Jia14,Jia15,Wan15,Tia16}, each nucleus in AME2012 with $20<$ $A$ $<$ $270$ is also considered equally.
Therefore, $c_{\rm sym,4}$ can be also treated as arithmetic average value of the $A$-dependent $a_{\rm{sym,4}}(A)$ (i.e., Eq.~(\ref{asym4})).
Following the discussion in Section~\ref{S-asym}, one can then extract information on $E_{\rm sym,4}(\rho_0)$ from $c_{\rm sym,4}$.
It is interesting to see from Fig.~\ref{F-asym4} that although the $a_{\rm{sym,4}}(A)$ exhibits $A$-dependence, it changes not so rapidly in the mass range of $A$ $<$ $270$. Such a feature makes it more reliable for us to estimate the value of $E_{\rm sym,4}(\rho_0)$ through the obtained constrains on $c_{\rm sym,4}$.

The $c_{\rm sym,4}$) can then be expressed as
\begin{eqnarray}
 c_{\rm sym,4} & \approx & \langle a_{\rm sym,4}\rangle = \sum\frac{1}{N_{\rm MN}}a_{\rm sym,4}(A)\nonumber\\
 & = & \Big(E_{\rm sym,4}(\rho_0)-\frac{L^2}{2K_0}\Big)\langle D_2^2(A)\rangle\label{asym4b},
\end{eqnarray}
where the second equation is obtained by using Eq.~(\ref{asym4}).
From Eq.~(\ref{asym4b}), one can then obtain $E_{\rm sym,4}(\rho_0)$ as
\begin{equation}
 E_{\rm sym,4}(\rho_0) = \frac{\langle a_{\rm sym,4}(A)\rangle}{\langle D_2^2(A)\rangle} + \frac{L^2}{2K_0}.
\label{Esym4}
\end{equation}
Since $\langle D_2^2(A)\rangle$ is a function of the symmetry volume-surface ratio $E_{\rm sym}(\rho_0)/\beta$, the $E_{\rm sym,4}(\rho_0)$ can then be determined by $L$, $K_0$, $\langle a_{\rm sym,4}\rangle$ and $E_{\rm sym}(\rho_0)/\beta$.
The error of $E_{\rm sym,4}(\rho_0)$ can be estimated through the error transfer formula
\begin{equation}
 \triangle_{E_{\rm sym,4}(\rho_0)} = \sqrt{\sum_{\rm i}\Big(\frac{\partial E_{\rm sym,4}(\rho_0)}{\partial x_{\rm i}}\Big)^2\triangle_{x_{\rm i}}^2},
\end{equation}
where $x_{\rm i}$ represents the quantities $L$, $K_0$, $\langle a_{\rm sym,4}\rangle$ and $E_{\rm sym}(\rho_0)/\beta$.

To determine the detailed value of $E_{\rm sym,4}(\rho_0)$ through Eq.~(\ref{Esym4}), an ambiguity appears since there have three extracted values of $c_{\rm sym,4}$, namely $3.28\pm0.50$ MeV and $8.47\pm0.49$ MeV extracted in Ref.~\cite{Jia14} and $8.33\pm1.21$ MeV extracted in Ref.~\cite{Tia16}.
Noting that the positive correlation between $\langle a_{\rm sym,4}\rangle$ and $E_{\rm sym,4}(\rho_0)$ in Eq.~(\ref{Esym4}), for a conservative estimate of $E_{\rm sym,4}(\rho_0)$~(by conservative here means the minimum value of $E_{\rm sym,4}(\rho_0)$), we use the smallest extracted value of $c_{\rm sym,4}$, namely $3.28\pm0.50$ MeV in Ref.~\cite{Jia14} (i.e., the gray band shown in Fig.~\ref{F-asym4}) to estimate the magnitude of $E_{\rm sym,4}(\rho_0)$.
Similar to the analysis of $a_{\rm sym}(A,Z)$ in section~\ref{S-asym},
in Ref.~\cite{Jia14}, all measured nuclei in AME2012 with 20 $<$ $A$ $<$ $270$ are considered equally.
Therefore, we have $N_{\rm MN}$ $=$ $2348$ and the sum in Eq.~(\ref{asym4b}) runs over these nuclei as well. Using $E_{\rm sym}(\rho_0)/\beta$ $=$ $2.4\pm0.4$~\cite{Dan03,Dan17}, then we find $\langle D_2^2(A)\rangle = 0.2$.
Combining the empirical constrains of $L$ $=$ $45.2\pm10.0~\rm MeV$~\cite{ZhangZ13} and $K_0$ $=$ $240\pm40~\rm MeV$~\cite{Shl06}, we then obtain $E_{\rm sym,4}(\rho_0) = 20.0\pm4.6$ MeV with the squared errors from $L$, $K_0$, $\langle a_{\rm sym,4}\rangle$ and $E_{\rm sym}(\rho_0)/\beta$ being $3.5$ MeV$^2$, $0.5$ MeV$^2$, $5.8$ MeV$^2$ and $11.3$ MeV$^2$, respectively.

We would like to point out that the detailed value of $E_{\rm sym,4}(\rho_0) = 20.0\pm4.6$ MeV estimated above relies on the empirical values of the lower-order parameters $L$ and $K_0$ of nuclear matter EOS as well as the extracted values of $E_{\rm sym}(\rho_0)/\beta$ and $\langle a_{\rm sym,4}\rangle$ from analyzing nuclear mass data.
For example, if we choose $L$ $=$ $58.7\pm28.1~\rm MeV$~\cite{Oer17} or $L$ $=$ $58.9\pm16.5~\rm MeV$~\cite{LiBA13}, the obtained $E_{\rm sym,4}(\rho_0)$ changes to $23.0\pm8.1~\rm MeV$ or $23.0\pm5.9~\rm MeV$, respectively.
Nevertheless, the choice of $L$ does not change the constrain on $E_{\rm sym,4}(\rho_0)$ much, and our results clearly indicate that a sizable positive $E_{\rm sym,4}(\rho_0)$ is necessary to describe the value of $c_{\rm sym,4}$ obtained from the double difference of the ``experimental'' symmetry energy extracted from the nuclear mass data.
Considering the fact that the majority of nuclear energy density functionals based on mean-field models give a fairly small magnitude of $E_{\rm sym,4}(\rho_0)$ with its value less than $2~\rm{MeV}$~\cite{ChenLW09,Cai12,Arg17,PuJ17} (Note: $E_{\rm sym,4}(\rho_0) = 2$ MeV leads to $c_{\rm sym,4} \approx \langle a_{\rm sym,4}\rangle = -0.45$ MeV),
the effects beyond the mean-field approximation, such as the short-range correlation effects, might be needed to explain such a sizable $E_{\rm sym,4}(\rho_0)$.
On the other hand, if $E_{\rm sym,4}(\rho_0)$ is indeed very small, then a novel mechanism is called for to explain the large value of $c_{\rm sym,4}$ from analyzing the data on nuclear masses.

\section{Conclusion and outlook}
By self-consistently considering the bulk, surface and Coulomb contributions to the nuclear mass, we have obtained an extended nuclear mass formula. In this mass formula, the symmetry energy $a_{\rm sym}(A,Z)$ and the fourth-order symmetry energy $a_{\rm{sym,4}}(A)$ of finite nuclei are related explicitly to the characteristic parameters of nuclear matter EOS.
In particular, using the recently extracted constant fourth-order symmetry energy coefficient $c_{\rm{sym,4}}$ from analyzing the double difference of the ``experimental'' symmetry energy extracted from nuclear masses, we have estimated for the first time a value of $E_{\rm sym,4}(\rho_0) = 20.0\pm4.6$ MeV for nuclear matter fourth-order symmetry energy at nuclear normal density $\rho_0$.

The significant value of $E_{\rm sym,4}(\rho_0) = 20.0\pm4.6$ MeV challenges the mean-field models which generally predict $E_{\rm sym,4}(\rho_0) \lesssim 2$ MeV. Therefore, it will be interesting to explore $E_{\rm sym,4}(\rho_0)$ within the framework of beyond the mean-field approximation (e.g., by considering the short range correlation effects). This would substantially improve our understanding on the properties of nuclear matter systems at extreme isospin, such as neutron stars.

\section*{Acknowledgements}
We would like to thank Hui Jiang, Kai-Jia Sun and Zhen Zhang for useful discussions.
This work was supported in part by
the Major State Basic Research Development Program (973 Program) in China under
Contract Nos. 2013CB834405 and 2015CB856904,
the National Natural Science Foundation of China under Grant Nos. 11625521, 11275125
and 11135011,
the Program for Professor of Special Appointment (Eastern Scholar) at Shanghai
Institutions of Higher Learning,
Key Laboratory for Particle Physics, Astrophysics and Cosmology, Ministry of
Education, China,
and the Science and Technology Commission of Shanghai Municipality (11DZ2260700).

\end{document}